\documentclass[journal=jctcce,manuscript=article]{achemso}
\usepackage{graphicx} 
\usepackage[version=4]{mhchem}
\usepackage{dcolumn}
\usepackage{bm}
\usepackage{amssymb,amsfonts}
\usepackage{amsmath}
\usepackage{subcaption}
\usepackage{units}
\usepackage{xcolor}
\usepackage{float}
\usepackage{multirow}
\allowdisplaybreaks

\newcommand{\ReSpect}{\textsc{ReSpect}}

\DeclareMathOperator{\Tr}{Tr}

\renewcommand{\vec}[1]{\boldsymbol{#1}}
\newcommand{\mat}[1]{\mathbf{#1}}

\newcommand{\dg}[1]{#1^{\dagger}}
\newcommand{\bra}[1]{\langle#1|}
\newcommand{\ket}[1]{|#1\rangle}

\newcommand{\highlight}[1]{{\color{black} #1}}

\title{Toward Accurate RIXS Spectra at Heavy Element Edges: A Relativistic Four-Component and Exact Two-Component TDDFT Approach}

\author{Lukas Konecny}
\affiliation{%
 Department of Inorganic Chemistry, Faculty of Natural Sciences, Comenius University, SK-84215 Bratislava, Slovakia
}%
\alsoaffiliation{%
 Hylleraas Centre for Quantum Molecular Sciences, Department of Chemistry, UiT The Arctic University of Norway, N-9037 Tromsø, Norway
}%
\alsoaffiliation{%
Max Planck Institute for the Structure and Dynamics of Matter, Center for Free Electron Laser Science, Luruper Chaussee 149, 22761 Hamburg, Germany
}%

\author{Muhammed A. Dada}%
\affiliation{%
 Department of Chemistry, The University of Memphis, Memphis, TN 38152, USA
}%

\author{Daniel R. Nascimento}%
\email{daniel.nascimento@memphis.edu}
\affiliation{%
 Department of Chemistry, The University of Memphis, Memphis, TN 38152, USA
}%

\author{Michal Repisky}
\email{michal.repisky@uit.no}
\affiliation{%
 Department of Physical and Theoretical Chemistry, Faculty of Natural Sciences, Comenius University, SK-84215 Bratislava, Slovakia
}%
\alsoaffiliation{%
 Hylleraas Centre for Quantum Molecular Sciences, Department of Chemistry, UiT The Arctic University of Norway, N-9037 Tromsø, Norway
}%

\begin{document}

\maketitle


\begin{abstract}
We present a relativistic time-dependent density functional theory (TDDFT) approach for the simulation of resonant inelastic X-ray scattering (RIXS) spectra, based on both a full four-component (4c) Dirac–Coulomb Hamiltonian and a modern atomic mean-field exact two-component (amfX2C) Hamiltonian model. The approach builds on the pseudo-wavefunction formalism and a core–valence separation scheme, enabling the efficient evaluation of couplings between two manifolds of excited states relative to a common ground state, as required for solving the Kramers–Heisenberg equation for RIXS. The relativistic formulation provides a variational description of scalar and spin–orbit relativistic effects, which are essential for accurately describing inner-shell excitations involved in RIXS processes. Its transformation to the 2c regime via the amfX2C Hamiltonian significantly reduces the computational cost while offering 4c-quality results by accounting for two-electron and exchange–correlation picture-change effects arising from the X2C transformation. In addition to two-dimensional RIXS maps, the methodology enables the direct evaluation of high-energy-resolution fluorescence detection (HERFD) and resonant X-ray emission spectra (RXES). Applications to 2p3d and 3d4f RIXS maps of selected ruthenium and uranium complexes demonstrate that the amfX2C approach reproduces reference 4c results and experimental spectra with high accuracy, capturing all key spectral features and providing reliable peak assignments.
\end{abstract}

\section{Introduction}

When Erwin Schr\"{o}dinger set to write his quantum mechanical wave equation~\cite{Schrodinger1926a,Schrodinger1926b}, he was aware of the theory of relativity and aimed to compose a relativistic equation~\cite{Moore1989}. However, he did not find the proper trick and proceeded with the non-relativistic version. His vision was later realized by Paul Dirac who successfully proposed a relativistic equation that described the fine structure of electronic spectra and explained the origin of spin, as well as predicted the existence of positrons~\cite{Dirac1928a,Dirac1928b}.
E. Schr\"{o}dinger and P. Dirac later shared the 1933 Nobel Prize in Physics \textit{for the discovery of new productive forms of atomic theory}~\cite{Nobel1933}.
Augmented for many-electron systems by the electron–electron interaction, involving either purely classical Coulomb repulsion or additional relativistic correction terms, such as the magnetic Gaunt~\cite{Gaunt1929} and retardation Breit~\cite{Breit1929} contributions, the equation found applications in quantum chemistry and has been used to explain phenomena referred to as relativistic effects, which are intractable within a nonrelativistic framework.~\cite{Pyykko1978,Pyykko1988}

The gold standard of relativistic quantum chemistry is the full four-component (4c) methodology based on the Dirac–Coulomb(–Breit) Hamiltonian~\cite{Pyykko1978,Pyykko1988,Liu2010,Saue2011}. At the same time, considerable effort has been devoted to the development of approximate relativistic two-component (2c) methods capable of reaching four-component accuracy. Among these, the exact two-component (X2C) Hamiltonian has gained wide popularity in recent years,~\cite{Jensen2005,Kutzelnigg2005,Liu2007,Ilias2007} as it reduces the 4c problem to a 2c one by means of simple algebraic transformations, without requiring explicit analytical expressions for higher-order relativistic corrections and/or property operators. In practice, several variants of the X2C Hamiltonian exist~\cite{Jensen2005,Kutzelnigg2005,Liu2007,Ilias2007,Sikkema2009,Peng2007,Liu2009,Peng2013,Filatov2013,Konecny2016,Goings2016,Liu2018,Knecht2022,Zhang2022,Ehrman2023}, whose accuracy varies and depends strongly on (i) the choice of the parent 4c Hamiltonian used to construct the 2c model and (ii) the treatment of picture-change-transformed two-electron and exchange–correlation (XC) integrals.~\cite{Knecht2022} In this context, Knecht and co-workers recently developed two simple yet computationally efficient and numerically accurate X2C models: the atomic mean-field (amfX2C) and the extended atomic mean-field (eamfX2C).~\cite{Knecht2022} Building on earlier work by Liu and Cheng~\cite{Liu2018}, these models incorporate full spin–orbit and scalar relativistic corrections from two-electron interactions (Coulomb, Coulomb–Gaunt, or Coulomb–Breit), account for the underlying XC functionals in Kohn--Sham DFT, and enable their extension to property calculations using both response theory~\cite{Konecny2023} and real-time approach~\cite{Moitra2023}. Here, we expand on these ideas to the generation of Resonant-Inelastic X-ray Scattering (RIXS) maps for molecular complexes containing heavy elements.

\begin{figure}
    \centering
    \includegraphics[width=0.50\linewidth]{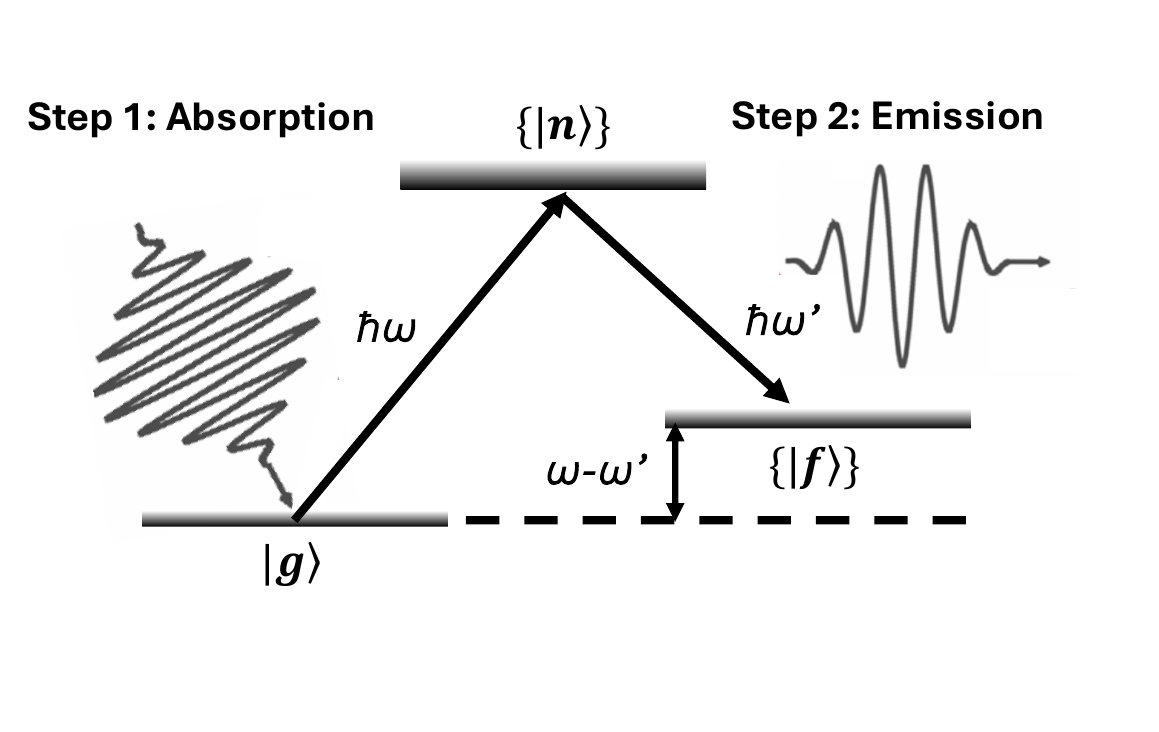}
    \vspace{-30pt}
    \caption{Schematic description of the RIXS process. A system in its ground state $\ket{g}$ is excited into a manifold of short-lived intermediate states $\{\ket{n}\}$ through the absorption of a resonant X-ray photon with energy $\omega$. The system then decays into one or more final states $\{\ket{f}\}$, emitting a photon of lower energy $\omega'$ and leaving the system in an excited state.}
    \label{fig:RIXS}
\end{figure}

RIXS is characterized by the absorption of an X-ray photon followed by the emission of photon with lower energy.~\cite{Gelmukhanov1999,Ament2011} The difference between the energies (or momenta) of the incoming and outgoing photons (the energy transfer) gives rise to spectral features that report on the electronic structure of valence or shallow core-level states inaccessible via direct one-photon absorption. As any other X-ray spectroscopy, RIXS offers high atomic selectivity, but with exceptional spectral resolution that is not limited by core-hole lifetimes~\cite{Gelmukhanov1999,Ament2011}. A schematic illustration of the RIXS process is shown in Figure~\ref{fig:RIXS}. For decades, RIXS studies were restricted to the condensed phase, but current advances in synchrotron radiation and free-electron laser facilities have enabled the extension of these experiments to gas- and solution-phase molecular systems.~\cite{Hennies:2010:193002,Nordgren:2013:0368,Kunnus:2013:16512,Pietzsch:2015:088302,Eckert:2017:6088,Ross:2018:5075,Hahn:2018:9515,Temperton:2019:074701,Fouda:2020:null,Biasin:2021:Revealing,Poulter2023,Larsen2024:28561} As this technique is increasingly applied to a broad range of molecular problems, reliable and computationally efficient electronic-structure methods become essential for the prediction and interpretation of complex spectral features.~\cite{Kvashnina2015Sensitivity, Pruessmann2022Opportunities, Zasimov2022HERFD, Schacherl2025Resonant}

In recent years, several theoretical approaches have been developed for the simulation of RIXS spectra in molecular systems. High-level wavefunction-based methods, such as damped response and equation-of-motion coupled-cluster theory~\cite{Faber:2019:520,Faber:2020:2642,Nanda:2020:2629,Nanda:2020:244118}, algebraic diagrammatic construction,~\cite{Rehn:2017:5552} multiconfigurational self-consistent field,~\cite{Josefsson2012AbComplexes, Polly2021Relativistic} matrix product states,~\cite{Lee:2023:7753} and configuration interaction,~\cite{Maganas:2017:11819} can describe RIXS processes in small molecules with high accuracy (see the review by Norman and Dreuw for a broader overview of electronic-structure approaches to X-ray spectroscopies~\cite{Norman2018}). Lower-cost, nonrelativistic approaches based on density functional theory (DFT) have also been developed with remarkable success.~\cite{Hanson-Heine2017Kohn-ShamSpectroscopy,Fouda2018SimulationWater,Besley2020DensitySpectroscopy,daCruz:2021:1835,Nascimento2021,Nascimento:2022:14680,Vitols2025Resonant}
Nonetheless, to enable the treatment of a broader range of systems, a relativistic approach is a necessity. This is particularly important for RIXS spectra at L and M edges, which require a multicomponent relativistic formalism to accurately describe excitations from p and d core orbitals split by spin–orbit interaction~\cite{Konecny2022}.

\highlight{Modern core-level time-dependent DFT (TDDFT) approaches can provide reliable excitation energies and transition intensities for a broad range of compounds, although certain classes of systems remain challenging. In open-shell transition-metal and actinide complexes, the interaction between the core hole and partially occupied valence shells can give rise to pronounced multiplet structure and dense manifolds of strongly mixed electronic states.~\cite{DeGroot2005, Butorin20203d} These features originate from many-electron effects, including strong core–valence Coulomb and exchange interactions, significant spin–orbit coupling, and multiconfigurational character associated with near-degenerate valence configurations. While TDDFT captures a substantial portion of the underlying physics through the exchange-correlation kernel, a fully quantitative description of such multiplet-rich spectra may require approaches that explicitly treat configuration interaction within the core-excited-state manifold, such as crystal-field or charge-transfer multiplet theories, multireference wavefunction methods, or related many-body techniques.~\cite{Haverkort2016Quanty, Zimmermann2018Quanty4rixs, Lundberg2019Multiconfig} Consequently, the choice of theoretical framework depends on the balance between computational efficiency, system size, and the degree of electronic correlation present in the system under investigation.}

In this work, we report a full four-component implementation of an approach based on TDDFT and the Dirac–Coulomb Hamiltonian for the simulation of two-dimensional resonant inelastic X-ray scattering (RIXS) maps, as well as their one-dimensional cuts corresponding to high-energy-resolution fluorescence detection (HERFD) and resonant X-ray emission spectra (RXES). The approach extends the earlier theoretical methodology of Nascimento \emph{et al.}~\cite{Nascimento2021} to the relativistic domain, thereby enabling a variational inclusion of scalar (SC) and spin–orbit (SO) relativistic effects that are particularly important for inner-shell orbitals involved in RIXS processes.
In addition, the relativistic approach is transformed into the two-component regime using the atomic mean-field exact two-component (amfX2C) Hamiltonian model, which accounts for SC and SO two-electron and exchange–correlation picture-change effects arising from the X2C transformation. We demonstrate that the amfX2C Hamiltonian reproduces the reference four-component results and predicts experimental RIXS features with remarkable accuracy, as illustrated by calculations of 2p3d and 3d4f RIXS maps for selected ruthenium and uranium complexes.

\section{Theory}

For isotropically averaged molecular orientations, RIXS spectra \highlight{in the soft X-ray region} are expressed in terms of the corresponding cross sections ($\sigma_\theta$) as two-dimensional functions of the incident ($\omega$) and emission ($\omega'$) frequencies using the \highlight{averaged} Kramers--Heisenberg formula (in atomic units)~\cite{Kramers1925,Gelmukhanov1999,Ament2011,Nascimento2021} \highlight{after neglecting interference between scattering channels},
\begin{equation}
\label{eq:RIXS}
\sigma_\theta (\omega',\omega)
=
\frac{\omega'}{\omega} \sum_{fn} \left| F_{fn} (\theta) \right|
\left[ \frac{(\omega_f-\omega_n)^{2} \omega_n^{2} \alpha^2}{(\omega - \omega_n)^2 + \Gamma_{n}^{2}/4} \right]
\Phi(\omega - \omega' - \omega_f,\gamma)
.
\end{equation}
Here, the summation runs over a manifold of intermediate states $\{\ket{n}\}$ and final states $\{\ket{f}\}$, where $\omega_{n}$ and $\omega_{f}$ denote the excitation energies to these states relative to the ground-state energy $\omega_{g}$. In addition, $\alpha$ is the fine structure constant in atomic units, $\Gamma_{n}$ is the Lorentzian broadening parameter associated with the intermediate state lifetime, and $\Phi$ is the lineshape function associated with the energy transfer $(\omega-\omega')$, with width $\gamma$ and center $\omega_f$. Finally, $\mathbf{F}(\theta)$ denotes the matrix of polarization-dependent transition amplitudes between intermediate and final states, whose values depend on the angle $\theta$ between the polarization of the incoming photon and the propagation direction of the emitted photon. Its matrix elements are defined as
\begin{equation}
\label{eq:Ftensor}
F_{fn}(\theta)
=
\frac{1}{15}\sum_{uv}^{x,y,z}\left[ \left( 2 - \frac{1}{2}\sin^2 \theta \right) (S_{fn}^{uv})^2
+ \left( \frac{3}{4}\sin^2 \theta - \frac{1}{2} \right) (S_{fn}^{uu} S_{fn}^{vv} + S_{fn}^{uv} S_{fn}^{vu}) \right]
,
\end{equation}
and involve the product of transition moments connecting the ground state to intermediate states and the intermediate states to final states,
\begin{equation}
\label{eq:Stensor}
S_{fn}^{uv}
= \bra{f} \dg{\hat{T}}_u \ket{n} \bra{n} \hat{T}_v \ket{0}.
\end{equation}
$\boldsymbol{\hat{T}}$ is a vector operator that couples a molecular system to an external electric field. In the present work, we adopt the long-wavelength approximation, in which $\boldsymbol{\hat{T}}$ reduces within the length gauge to the electric dipole moment operator. It is worth noting that the long-wavelength approximation is not always valid within the X-ray regime; thus, the inclusion of higher-order multipoles in the deﬁnition of $\boldsymbol{\hat{T}}$ may become necessary.~\cite{List2015, List2017, List2020}

Within the linear response TDDFT (LR-TDDFT), the transition moments between ground and excited states may easily be extracted from the
solutions of LR-TDDFT
\begin{equation}
\label{eq:0-to-n}
\bra{n} \hat{T}_v \ket{0}
=
\sum_{ai} \left(
T_{ia}^{v} X_{ai}^n
+
T_{ai}^{v} Y_{ai}^n
\right)
\end{equation}
where $T_{ia}^{v}$ is a matrix element of the electric dipole moment operator $\hat{T}_v$ in the reference molecular orbital basis $\{\varphi\}$,
\begin{equation}
  \label{eq:el-dipole}
  T_{ia}^{v}
  =
  \bra{\varphi_i} \hat{T}_{v} \ket{\varphi_{a}}
  =
  -\bra{\varphi_i} \hat{r}_{v}-R_{g,v} \ket{\varphi_{a}}.
\end{equation}
Here, $i, j, ...$ index occupied molecular orbitals (MOs), $a, b, ...$ index unoccupied MOs, $\boldsymbol{\hat{r}}$ denotes the electron position operator defined with respect to a gauge origin $\boldsymbol{R}_{g}$, which is taken in this work to be the nuclear charge center of the molecule. The matrix elements $X_{ai}^n$ and $Y_{ai}^n$ in Eq.~\eqref{eq:0-to-n} correspond to the solutions of the eigenvalue LR-TDDFT equation
\begin{equation}
  \label{eq:LR-TDDFT}
  \begin{pmatrix}
   \mathbf{A}      &  \mathbf{B} \\
  -\mathbf{B}^\ast & -\mathbf{A}^{\!\ast}
  \end{pmatrix}
  \begin{pmatrix}
  \mathbf{X}^{n} \\
  \mathbf{Y}^{n}
  \end{pmatrix}
  =
  \omega_{n}
  \begin{pmatrix}
  \mathbf{X}^{n} \\
  \mathbf{Y}^{n}
  \end{pmatrix}
\end{equation}
where $\omega_n$ is the vertical excitation energy from the reference electronic ground state to the $n$th excited electronic state, and $(\mathbf{X}^{n}~\mathbf{Y}^{n})^{\mathrm{T}}$ is the corresponding transition vector with nonzero elements $(X^{n}_{ai}~Y^{n}_{ai})^{\mathrm{T}}$ in the virtual–occupied block. Essential details regarding the evaluation of the matrices $\mathbf{A}$ and $\mathbf{B}$ within the two- and four-component frameworks are discussed below.~\cite{Komorovsky2019,Konecny2019,Konecny2023}

To calculate operator matrix elements between two excited states, as needed in Eq.~\eqref{eq:Stensor}, would typically require second-order response theory.~\cite{Dalgaard1982Quadratic}
Unfortunately, the quadratic response formalism is associated with a substantially higher computational cost and, within approximate frameworks such as TDDFT, may exhibit spurious poles when the energy difference between two excited states coincides with the excitation energy of another state.~\cite{Parker2016Unphysical, Parker2018Quadratic, Bowman2017Excited}
As an alternative, one may extract approximate excited-state couplings directly from LR-TDDFT amplitudes using the so-called \emph{pseudo-wavefunction} formalism.~\cite{Ou2015First, Ou2015Derivative, Zhang2015analytic, Alguire2015, Parker2018Quadratic, Sheng2020Excited}
The same approach has previously been considered by Nascimento et al. in the context of resonant inelastic X-ray scattering (RIXS) calculations.~\cite{Nascimento2021}

In the TDDFT pseudo-wavefunction formalism, the ansatz for an excited-state wavefunction $\ket{k}$ is
\begin{equation}
   \ket{k}
   \equiv
   \left(
      \hat{X}^{k} + \hat{X}^{k}\hat{X}^{l}\hat{Y}^{l}
   \right) \ket{\mathrm{KS}}
\end{equation}
where $\left| \mathrm{KS} \right\rangle$ is the reference Kohn--Sham ground-state Slater determinant, and the excitation operators take the form
\begin{equation}
   \hat{X}^{l} = \sum_{ai} X_{ai}^{l} \hat{a}_{a}^{\dagger} \hat{a}_{i}
   ;\qquad
   \hat{Y}^{l} = \sum_{ai} Y_{ai}^{l} \hat{a}_{a}^{\dagger} \hat{a}_{i}.
\end{equation}
The amplitudes $\mathbf{X}^{l}$ and $\mathbf{Y}^{l}$ are the solutions of the LR-TDDFT equation for the $l$th excited electronic state, whereas $\dg{\hat{a}}_a$, $\hat{a}_i$ are single-particle creation and annihilation operators, respectively.

For spectroscopic properties pertaining to the core region, sufficiently accurate results can be obtained~\cite{Dada2026Quantifying} by approximating the
excited pseudo-wavefunction as linear combinations of singly substituted Slater determinants with the expansion coefficients provided by eigenvectors obtained as solutions of Eq.~\eqref{eq:LR-TDDFT} within the Tamm--Dancoff approximation~\cite{Hirata1999}, \emph{i.e.} by approximating $\mathbf{B}=\mathbf{0}$,
yielding
\begin{equation}
\ket{k}
\approx
\sum_{ai} X^{k}_{ai} \dg{\hat{a}}_a\hat{a}_i \ket{\mathrm{KS}}
.
\end{equation}
After expressing both the intermediate and final state in this way, an operator matrix element between them can be calculated as
\begin{equation}
\left\langle f \right| \dg{\hat{T}}_u \left| n \right\rangle
=
\sum_{ai} X^{f}_{ai} \left( \sum_{b} \tilde{T}_{ab}^u X^{n}_{bi} - \sum_{j} X^{n}_{aj} \tilde{T}_{ji}^u \right)
=
-\sum_{ai} X^{f}_{ai} \sum_{j} X^{n}_{aj} \tilde{T}_{ji}^u,
\end{equation}
where the last equality holds when core--valence separation is used in the TDDFT calculation and the intermediate
and final excited states correspond to excitations from different occupied orbitals, \emph{i.e.} $i$ and $j$ span
a different subset of occupied orbitals.

The TDDFT RIXS working equations have so far been formulated in a general manner, independent of the one-, two-, or four-component framework. To render the presentation self-contained, we summarize in this paragraph the terms appearing in Eq.~\eqref{eq:LR-TDDFT} that are specific to the four-component (4c) framework. Now, assuming implicit summation over repeated indices and an orthonormal atomic orbital (AO) basis indexed by the Greek letters $\mu, \nu, \kappa, \lambda$, the matrices $\mathbf{A}$ and $\mathbf{B}$ take the following form:~\cite{Komorovsky2019,Konecny2019}
\begin{equation}
\begin{alignedat}{2}
 \label{eq:4cAB}
 A_{ai,bj}^{\text{4c}}
 &=
 (\varepsilon_a^{\text{4c}} - \varepsilon_i^{\text{4c}})\delta_{ab}\delta_{ij}
 &~+~
 (g_{\mu\nu,\kappa\lambda}^{\text{4c}} - k_{\mu\nu,\kappa\lambda}^{\text{4c}})
 C^{\text{4c}^\ast}_{\mu a} C^{\text{4c}}_{\nu i} C^{\text{4c}^\ast}_{\kappa j} C^{\text{4c}}_{\lambda b},
 \\
 B_{ai,bj}^{\text{4c}}
 &=&
 (g_{\mu\nu,\kappa\lambda}^{\text{4c}} - k_{\mu\nu,\kappa\lambda}^{\text{4c}})
 C^{\text{4c}^\ast}_{\mu a} C^{\text{4c}}_{\nu i} C^{\text{4c}^\ast}_{\kappa b} C^{\text{4c}}_{\lambda j},
\end{alignedat}
\end{equation}
where the molecular orbital coefficients $C^{\text{4c}}_{\mu p}$ and one-electron energies
$\varepsilon^{\text{4c}}_p$ are solutions of the Dirac--Kohn--Sham equations (the summation over
index $p$ is not implied)
\begin{equation}
  \left[
    h^{\text{D}}_{\mu\nu} + G^{\text{4c}}_{\mu\nu} + V^{\text{4c}}_{\text{xc},\mu\nu}
  \right]
  C^{\text{4c}}_{\nu p}
  =
  C^{\text{4c}}_{\mu p} \varepsilon^{\text{4c}}_p.
\end{equation}
The 4c Fock matrix on the left-hand side, $\mat{F}^{\text{4c}}=\mat{h}^{\text{D}} + \mat{G}^{\text{4c}} + \mat{V}^{\text{4c}}_{\text{xc}}$, within the Dirac--Coulomb Hamiltonian framework assumed in this work, consists of the Dirac one-electron, Coulomb two-electron, and noncollinear exchange--correlation (xc) contributions:~\cite{Repisky2020,Knecht2022,Repisky2025}
\begin{subequations}
\begin{align}
  \label{eq:h4c}
  h^{\text{D}}_{\mu\nu}
  &=
  \int
  X^{\text{4c}^{\dagger}}_{\mu}(\vec{r})
  [c(\vec{\alpha} \cdot \vec{p}) + c^2 (\beta-I_{4}) + \hat{V}^{\text{ne}}(\vec{r})]
  X^{\text{4c}}_{\nu}(\vec{r}) d^{3}\vec{r}
\\
  \label{eq:G4c}
  G^{\text{4c}}_{\mu\nu}
  &=
  g^{\text{4c}}_{\mu\nu,\kappa\lambda}
  D^{\text{4c}}_{\lambda\kappa}
  \quad
  \begin{cases}
    D^{\text{4c}}_{\lambda\kappa} = C^{\text{4c}}_{\lambda i} C^{\text{4c}^{\ast}}_{\kappa i}
    \\[0.2cm]
    g^{\text{4c}}_{\mu\nu,\kappa\lambda}
    =
    [\Omega^{\text{4c}}_{\mu\nu}|\Omega^{\text{4c}}_{\kappa\lambda}]
    -
    \xi[\Omega^{\text{4c}}_{\mu\lambda}|\Omega^{\text{4c}}_{\kappa\nu}]
    \\[0.2cm]
    [\Omega^{\text{4c}}_{\mu\nu}|\Omega^{\text{4c}}_{\kappa\lambda}]
    =
    \iint
    \Omega^{\text{4c}}_{\mu\nu}(\vec{r}_{1})
    \,r_{12}^{-1}\,
    \Omega^{\text{4c}}_{\kappa\lambda}(\vec{r}_{2})
    d^{3}\vec{r}_{1}d^{3}\vec{r}_{2}
    \\[0.3cm]
  \end{cases}
\\
  \label{eq:XC4c}
  V^{\text{4c}}_{\text{xc},\mu\nu}
  &=
  \int
  v^{\text{xc}}_k[\vec{\rho}^{\text{4c}}](\vec{r})
  \Omega^{\text{4c}}_{k,\mu\nu}(\vec{r}) d^{3}\vec{r}
  \quad
  \begin{cases}
    \Omega^{\text{4c}}_{k,\mu\nu}
    =
    X^{\text{4c}^{\dagger}}_{\mu} \Sigma_k X^{\text{4c}}_{\nu}
    \\[0.2cm]
    \rho_k^{\text{4c}}
    =
    \Tr
    \Big\{
    \mat{\Omega}_k^{\text{4c}}
    \mat{D}^{\text{4c}}
    \Big\}
    \\[0.2cm]
    v^{\text{xc}}_k[\vec{\rho}^{\text{4c}}]
    =
    \frac{\partial \varepsilon^{\text{xc}}}{\partial \rho_k^{\text{4c}}}
    -\left(
    \vec{\nabla}\cdot\frac{\partial\varepsilon^{\text{xc}}}{\vec{\nabla}\rho_k^{\text{4c}}} \right)
    \\[0.3cm]
  \end{cases}
\end{align}
\end{subequations}
In Eq.~\eqref{eq:h4c}, the Dirac Hamiltonian is represented in a finite 4c restricted kinetically balanced basis
$\{X^{\text{4c}}\}$~\cite{Stanton1984} and involves the canonical momentum operator $\vec{p}$ and the nuclear–electron electrostatic potential operator $\hat{V}^{\text{ne}}$, where the nuclear charge distribution is modeled by a spherical Gaussian function~\cite{Visscher1997}. Here, $c \simeq \unit[137]{a.u.}$ denotes the speed of light, and $\vec{\alpha}$ and $\beta$ refer to the Dirac matrices in the standard representation, with $\beta$ shifted by the $4 \times 4$ identity matrix $I_{4}$. In Eq.~\eqref{eq:G4c}, the two-electron contribution involves the one-electron density matrix $\mat{D}^{\text{4c}}$, contracted with the 4c electron repulsion integrals $\mat{g}^{\text{4c}}$, which are weighted in the exchange part by a scalar parameter $\xi$ to accommodate hybrid DFT functions.
Finally, in Eq.~\eqref{eq:XC4c}, $v_k^{\text{xc}}$ denotes the \emph{noncollinear} exchange–correlation potential, parametrized within the generalized gradient approximation by the scalar electronic charge density $(\rho^{\text{4c}}_{k=0})$ and its gradient $(\vec{\nabla}\rho^{\text{4c}}_{k=0})$, as well as by the three components of the electronic spin density $(\rho^{\text{4c}}_{k=1,2,3})$ and their gradients $(\vec{\nabla}\rho^{\text{4c}}_{k=1,2,3})$. All densities are defined in terms of the corresponding 4c charge ($\Sigma_0$) and spin operators ($\Sigma_{k=1,2,3}$), and are crucial for the correct description of spin-density distribution in relativistic theory.
For details on the noncollinear formulation of XC potential and kernel in our \ReSpect{} program interested readers are referred to Refs.~\citenum{Komorovsky2019,Repisky2020}.

To obtain the TDDFT RIXS working equations within the exact two-component (X2C) approach, let us first recall that the central idea of the X2C approach is the transformation of the full four-component Fock matrix of size
$\mathbb{C}^{4n\times4n}$ into its block-diagonal form using a unitary decoupling matrix
$\mat{U}\in\mathbb{C}^{4n\times4n}$~\cite{Jensen2005,Kutzelnigg2005,Liu2007,Ilias2007}:
\begin{eqnarray}
      \mat{F}^{\text{4c}}
      \rightarrow
      \tilde{\mat{F}}^{\text{4c}}
      =
      \mat{U}^{\dagger}\mat{F}^{\text{4c}}\mat{U}
      =
      \begin{pmatrix}
         \tilde{\mat{F}}_{11} & \tilde{\mat{F}}_{12} & \mat{0}              & \mat{0} \\
         \tilde{\mat{F}}_{21} & \tilde{\mat{F}}_{22} & \mat{0}              & \mat{0} \\
         \mat{0}              & \mat{0}              & \tilde{\mat{F}}_{33} & \tilde{\mat{F}}_{34} \\
         \mat{0}              & \mat{0}              & \tilde{\mat{F}}_{43} & \tilde{\mat{F}}_{44} \\
      \end{pmatrix}
      .
\end{eqnarray}
Thanks to the unitary property of $\mat{U}$, all eigenvalues of the parent 4c problem can be reproduced to computer precision by solving two sets of uncoupled Fock equations, each with half the dimension of the original problem. By disregarding the set associated with negative-energy solutions, we are left with the X2C--Kohn--Sham equations
for the positive-energy solutions (++), \emph{i.e.}
\begin{eqnarray}
    \tilde{\mat{F}}^{\text{2c}} \tilde{\mat{C}}^{\text{2c}}
    =
    \tilde{\mat{C}}^{\text{2c}} \epsilon^{\text{2c}}
    \quad
    \begin{cases}
    \tilde{\mat{C}}^{\text{2c}}
    =
    \big[ \mat{U}^{\dagger}\mat{C}^{\text{4c}} \big]^{++}
    ~=
    \begin{pmatrix}
        \tilde{\mat{C}}_{11} & \tilde{\mat{C}}_{12} \\
        \tilde{\mat{C}}_{21} & \tilde{\mat{C}}_{22} \\
    \end{pmatrix}
    \\
    \tilde{\mat{F}}^{\text{2c}}
    =
    \big[ \mat{U}^{\dagger}\mat{F}^{\text{4c}}\mat{U} \big]^{++}
    =
    \begin{pmatrix}
        \tilde{\mat{F}}_{11} & \tilde{\mat{F}}_{12} \\
        \tilde{\mat{F}}_{21} & \tilde{\mat{F}}_{22} \\
    \end{pmatrix}
    \end{cases}
\end{eqnarray}
Note that all two-component quantities undergoing the \emph{exact} two-component transformation are marked with a tilde. Given that the positive-energy 4c molecular orbitals coefficients can be expressed in terms of their 2c counterparts,
$\mat{C}^{\text{4c}} = \mat{U}\tilde{\mat{C}}^{\text{2c}}$, Eq.~\eqref{eq:4cAB} can be formulated entirely in terms of 2c quantities
\begin{equation}
\begin{alignedat}{2}
 \label{eq:2cAB}
 A_{ai,bj}^{\text{2c}}
 &=
 (\varepsilon_a^{\text{2c}} - \varepsilon_i^{\text{2c}})\delta_{ab}\delta_{ij}
 &~+~
 (\tilde{g}_{\mu\nu,\kappa\lambda}^{\text{2c}} - \tilde{k}_{\mu\nu,\kappa\lambda}^{\text{2c}})
 \tilde{C}^{\text{2c}^\ast}_{\mu a} \tilde{C}^{\text{2c}}_{\nu i} \tilde{C}^{\text{2c}^\ast}_{\kappa j} \tilde{C}^{\text{2c}}_{\lambda b},
 \\
 B_{ai,bj}^{\text{2c}}
 &=&
 (\tilde{g}_{\mu\nu,\kappa\lambda}^{\text{2c}} - \tilde{k}_{\mu\nu,\kappa\lambda}^{\text{2c}})
 \tilde{C}^{\text{2c}^\ast}_{\mu a} \tilde{C}^{\text{2c}}_{\nu i} \tilde{C}^{\text{2c}^\ast}_{\kappa b} \tilde{C}^{\text{2c}}_{\lambda j}.
\end{alignedat}
\end{equation}
Here, $\tilde{\mat{g}}^{\text{2c}}$ and $\tilde{\mat{k}}^{\text{2c}}$ are the X2C-transformed two-electron and exchange–correlation kernels~\cite{Knecht2022,Konecny2023}. In general, their evaluation requires a costly four-index transformation, making 2c calculations more demanding than their 4c counterparts. In the current implementation, we approximate the transformed kernels by the untransformed ones, \emph{i.e.}, $\tilde{\mat{g}}^{\text{2c}} \approx \mat{g}^{\text{2c}}$, and similarly $\tilde{\mat{k}}^{\text{2c}} \approx \mat{k}^{\text{2c}}$, where the untransformed quantities are evaluated in a conventional large-component basis. As shown for XAS, this approximation results in negligibly small errors and yields 2c results of 4c quality at a fraction of the 4c computational cost.~\cite{Konecny2023} Finally, the 2c molecular orbital coefficients $\tilde{C}^{\text{2c}}_{\mu p}$ and one-electron energies $\varepsilon^{\text{2c}}_{p}$ in Eq.~\eqref{eq:2cAB} are obtained as solutions of the X2C--Kohn--Sham equations using the recently developed amfX2C Hamiltonian model:~\cite{Knecht2022,Repisky2025}
\begin{equation}
  \Big[
      \underbrace{
      \mat{\tilde{h}}^{\text{2c}}
      +
      \mat{G}^{\text{2c}}
      +
      \Delta\mat{G}^{\text{2c}}_{\bigoplus}
      +
      \mat{V}^{\text{2c}}_{\text{xc}}
      +
      \Delta\mat{V}^{\text{2c}}_{\text{xc}\bigoplus}
      }_{\mat{F}^{\text{amfX2C}}}
  \Big]_{\mu\nu}
  \tilde{C}^{\text{2c}}_{\nu p}
  =
  \tilde{C}^{\text{2c}}_{\mu p} \varepsilon^{\text{2c}}_p.
\end{equation}
Here, $\Delta\mat{G}^{\text{2c}}_{\bigoplus}$ and $\Delta\mat{V}^{\text{2c}}_{\text{xc}\bigoplus}$ are approximate two-electron and XC picture-change correction terms assembled from atomic 4c calculations prior to the molecular 2c SCF procedure~\cite{Knecht2022}. In contrast, $\mat{G}^{\text{2c}}$ and $\mat{V}^{\text{2c}}_{\text{xc}}$ are updated throughout the SCF iterations and require only conventional nonrelativistic two-electron and XC integrals. Based on our experience, the computational cost of amfX2C SCF calculations does not exceed a factor of four compared to nonrelativistic calculations, while achieving an energy accuracy of approximately 10 $\mu$Hartree per atom relative to the full 4c treatment~\cite{Knecht2022}.
\highlight{This level of accuracy also carries over to all quantities entering the response-theory equations, as evidenced by the XAS results of Ref.~\citenum{Konecny2023} and the RIXS results reported in the present work.}

\section{Computational Details}

The developed methodology for the calculations of RIXS spectra was benchmarked for the \ce{L3}-edge
of ruthenium complex \ce{[Ru^{II}(CN)6]^{4-}}, and \ce{M4}- and \ce{M5}-edges of uranium(IV) hexahalide
complexes, \ce{[U^{IV}X6]^{2-}} (X = F, Cl, or Br), all considered in a closed-shell electronic configuration. All considered systems possess octahedral symmetry with bond lengths
$r(\textrm{U-F}) = \unit[2.17]{\AA}$,
$r(\textrm{U-Cl}) = \unit[2.62]{\AA}$,
and $r(\textrm{U-Br}) = \unit[2.75]{\AA}$,
taken from reference experimental work~\cite{Burrow2024},
and $r(\textrm{Ru-C}) = \unit[2.077]{\AA}$ and $r(\textrm{C-N}) = \unit[1.173]{\AA}$
taken from our previous theoretical works~\cite{Biasin:2021:Revealing, Nascimento2021}.
\highlight{The U(IV) complexes studied here formally possess a $5f^2$ configuration and are therefore weakly paramagnetic. While the closed-shell TDDFT treatment adopted in this work does not explicitly account for possible multiconfigurational effects, the substantial ligand-field splitting of the uranium $5f$ orbitals reported by Burrow \textit{et al.},\cite{Burrow2024}, together with the overall agreement between theory and experiment, suggests that the closed-shell treatment provides a reasonable approximation.}

All RIXS spectra were calculated using the eigenvalue linear response TDDFT library~\cite{Komorovsky2019,Konecny2019} of our relativistic spectroscopy DFT program \ReSpect{}~\cite{Repisky2020,Repisky2025}.
All calculations were performed with uncontracted bases comprising of Dyall's VDZ basis for uranium~\cite{Dyall2007-5f} and ruthenium~\cite{Dyall2007-4d}, and Dunning's aug-cc-pVDZ for all remaining elements~\cite{Dunning1989,Kendall1992,Woon1993}. \highlight{The basis-set choice was guided by our previous benchmark study of 4c TDDFT X-ray absorption spectra at the L$_{2,3}$- and M$_{4,5}$-edges of closed-shell transition-metal and actinide compounds covering a range of central atoms, ligands, and oxidation states~\cite{Konecny2022}.} The PBE0 exchange--correlation (XC) functional~\cite{Slater1951,Perdew1996,Perdew1997,Adamo1999} was used in the calculation
pertaining the \ce{[Ru^{II}(CN)6]^{4-}} complex, while a modified version of this functional
with 60\% admixture of Hartree--Fock exchange (PBE0-60HF) was used for the uranium(IV) hexahalide anions
according to the observation that the functional reduces the self-interaction error of XC functions and provides improved absorption peak positions~\cite{Konecny2022}. The numerical integration of the noncollinear exchange--correlation potential and kernel was done with an adaptive molecular grid of medium size (program default), and atomic nuclei of finite size were approximated by a Gaussian charge distribution model~\cite{Visscher1997}. \highlight{To accelerate the most demanding part of 4c calculations, namely the evaluation of four-center electron-repulsion integrals (ERIs) over small-component basis functions ($\chi^{\text{S}}$), we employed an atom-pair approximation in which these integrals were neglected unless the bra and ket basis pairs shared the same atomic center, i.e., $[ \chi_A^\mathrm{S} \chi_B^\mathrm{S} | \chi_C^\mathrm{S} \chi_D^\mathrm{S} ] \delta_{AB}\delta_{CD}$. Notably, this approximation is fully consistent with the amfX2C approach, where the atomic 4c ERIs are themselves evaluated exactly, without any approximation. In our experience, the atom-pair approximation introduces negligible errors in the 4c total energy $(<10^{-5}E_h)$ while reducing the computational cost by a factor of 2–3.}

To calculate the RIXS spectra, two TDDFT calculations within the Tamm--Dancoff approximation~\cite{Hirata1999} (TDA) were performed to obtain intermediate and final state excitation energies and transition vectors as discussed in
the Theory section. The TDDFT calculations employed core-valence separation by considering excitations only
from the relevant inner core and upper core orbitals, specifically Ru 2\ce{p_{3/2}} (inner core) and all 3d (outer core), and U 3\ce{d_{3/2}} (inner core for \ce{M4}-edge), 3\ce{d_{5/2}} (inner core for \ce{M5}-edge), and all 4f (outer core), while considering excitations to all virtual orbitals.
\highlight{In all TDDFT calculations, the first 3000 excitations were included in the evaluation of the RIXS spectra to ensure adequate coverage of the experimentally relevant spectral ranges given the high-density-of-state excitation manifolds.}
In Eq.~\eqref{eq:RIXS}, the lifetime broadening of the individual intermediate states, $\Gamma_n$, is approximated by a single empirical parameter, $\Gamma$, corresponding to the full width at half-maximum (FWHM) of the spectral lines. In addition, the energy-transfer lineshape function $\Phi$ is chosen to be a Gaussian function \highlight{modeling the instrumental resolution}, defined as
\begin{equation}
\Phi(\omega - \omega' - \omega_f,\gamma)
=
e^{-\frac{(\omega-\omega'-\omega_f)^2}{2\sigma^2}}; \qquad \sigma = \frac{\gamma}{2 \sqrt{2\ln 2}}
.
\end{equation}
Additionally, the resonant X-ray emission spectra (RXES) were then obtained from the RIXS 2D maps as vertical cuts taken at constant absorption energies determined as maxima of XAS spectra evaluated from the inner core TDDFT calculations. Furthermore, high energy resolution fluorescence detection (HERFD) spectra were obtained as cuts
of the RIXS 2D maps taken at constant emission energies determined from the RXES spectra. The values of absorption and emission maxima for the uranium(IV) hexalide anions are summarized in Table~\ref{tab:UX6:AbsEmiE}.
Similarly, constant energy transfer (CET) cuts of the RIXS maps were taken at constant energy transfer $\omega-\omega'$, where the constant absorption and emission energies were again obtained from XAS and RXES spectra, respectively.

\begin{table}
    \centering
    \begin{tabular}{lccccc}
    \hline
    \hline
     \multirow{2}{*}{System} & \multicolumn{2}{c}{\ce{M5}-edge} & & \multicolumn{2}{c}{\ce{M4}-edge}  \\
     \cline{2-3} \cline{5-6}
                          & absorption   & emission    & &  absorption   & emission    \\
     \hline
     \ce{[UF6]^{2-}}      &  3549.7      & 3171.5      & &  3726.6      &  3338.4     \\
     \ce{[UCl6]^{2-}}     &  3549.3      & 3170.9      & &  3727.5      &  3339.0     \\
     \ce{[UBr6]^{2-}}     &  3549.2      & 3170.9      & &  3727.5      &  3339.0     \\
     \hline
     \hline
    \end{tabular}
    \caption{Absorption and emission maxima (in eV) used in our simulations of \ce{[UX6]^{2-}}. Absorption maxima were determined from X-ray absorption spectra computed with TDDFT (TDA, amfX2C, PBE0-60HF), while the emission
    maxima were obtained from resonant X-ray emission spectra evaluated as cuts of the RIXS spectra
    at the absorption maxima.}
    \label{tab:UX6:AbsEmiE}
\end{table}

\section{Results}

The theoretical results were benchmarked against recent 2p3d and 3d4f RIXS experiments on selected heavy-element systems. The former involves the ruthenium \ce{L3}-edge in \ce{[Ru^{II}(CN)6]^{4-}},~\cite{Poulter2023} whereas the latter concerns the uranium \ce{M4}- and \ce{M5}-edges in hexahalide complexes, \ce{[U^{IV}X6]^{2-}} (X = F, Cl, or Br).~\cite{Burrow2024} In these systems, characteristic RIXS spectral features such as positions, lineshapes, and intensities are governed by the spin–orbit coupling of 2p, 3d, and 4f orbitals. Consequently, RIXS spectroscopy provides a highly sensitive probe of the quality of developed theoretical methods.

\begin{figure}[H]
    \centering
    \includegraphics[width=1.0\linewidth]{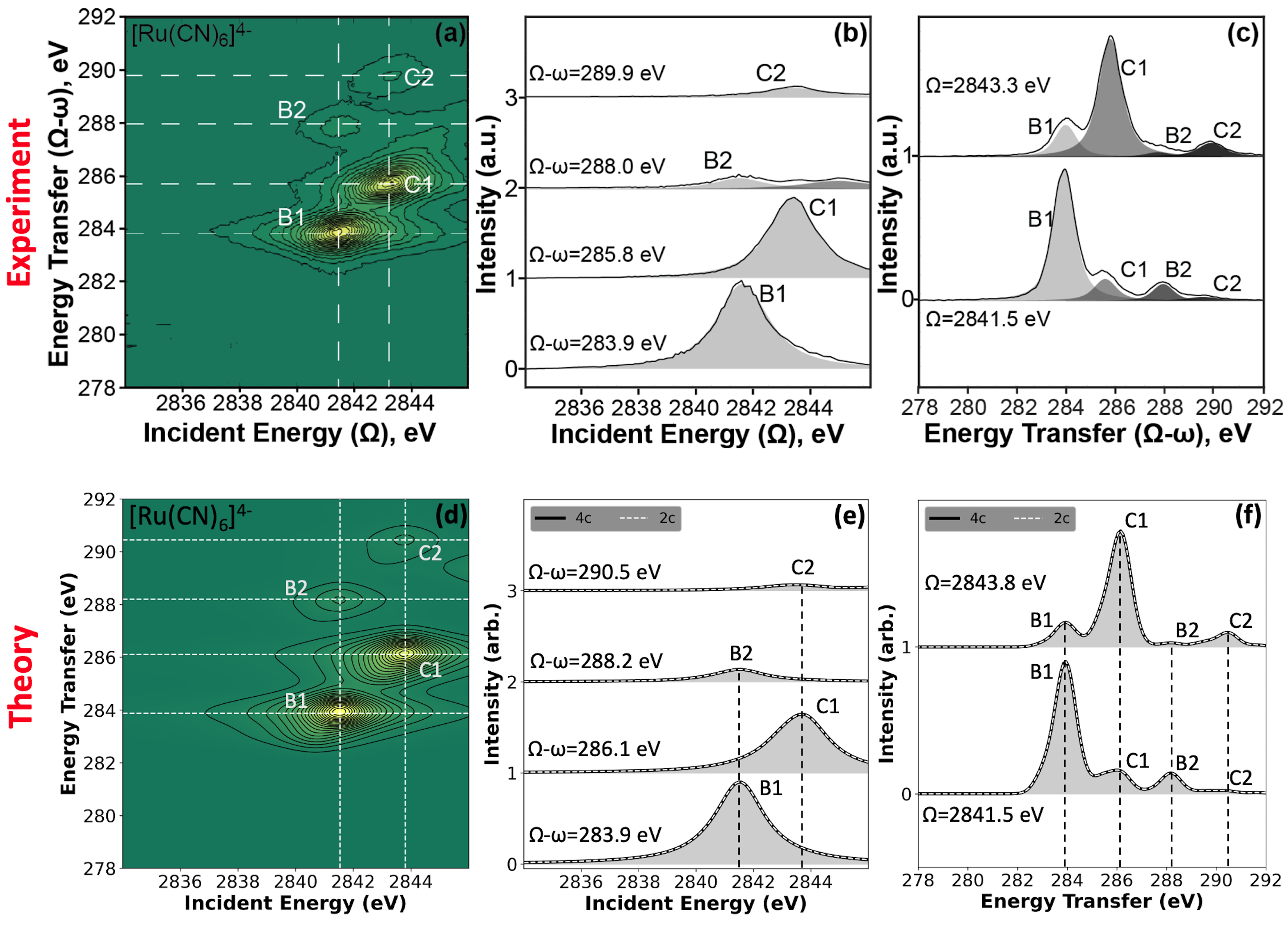}
    \caption{Ru \ce{L3}-edge 2p3d RIXS maps and the corresponding constant energy transfer (CET) and constant incident energy (CIE) cuts of \ce{[Ru(CN)6]^{4-}}, obtained experimentally (a–c) from Ref.~\citenum{Poulter2023} and theoretically (d–f) in this work. Theoretical spectra were computed at the relativistic four-component (4c) and amfX2C (2c) TDDFT level using the PBE0 functional, applying an incident-energy shift of \unit[35.69]{eV}, an energy-transfer shift of \unit[6.84]{eV}, and broadening parameters $\Gamma = \unit[2.1]{eV}$ and $\gamma = \unit[0.9]{eV}$ discussed in Eq.~\eqref{eq:RIXS}. Both RIXS maps are normalized to the maximum of the B1 peak, displayed using 20 evenly spaced contour levels, and marked with white dashed lines indicating the positions of the CET and CIE cuts. Experimental peaks in the 2D RIXS map (a) are located at (in eV) B1 (2841.5, 283.9), B2 (2841.5, 288.0), C1 (2843.3, 285.8), and C2 (2843.3, 289.9). Theoretical peaks in the 2D RIXS map (d) are located at (in eV) B1 (2841.5, 283.9), B2 (2841.5, 288.2), C1 (2843.8, 286.1), and C2 (2843.8, 290.5). The experimental (b) and theoretical (e) CET cuts are taken through the maxima of the corresponding B1, C1, B2, and C2 peaks. Similarly, the experimental (c) and theoretical (f) CIE cuts are taken through the maxima of the corresponding B1 and C1 peaks. Note that the notation for incident energy and energy transfer ($\omega$, $\omega-\omega'$) used throughout this paper was adapted in this figure to match the notation ($\Omega$, $\Omega-\omega$) of Ref.~\citenum{Poulter2023}.}
    \label{fig:RIXS:RuCN6}
\end{figure}

\textbf{\ce{[Ru^{II}(CN)6]^{4-}}:} The experimental and computed Ru \ce{L3} 2p3d RIXS maps of \ce{[Ru(CN)6]^{4-}} are shown in Figures \ref{fig:RIXS:RuCN6}a and \ref{fig:RIXS:RuCN6}d. Both spectra exhibit two distinct edge features, labeled B and C~\cite{vanKuiken2012Probing, vanKuiken2013Simulating,Poulter2023}. The B features, consisting of the B1 and B2 peaks along a constant incident energy cut (CIE) of 2841.5 eV, arise predominantly from resonant transitions from the core 2\ce{p_{3/2}} orbitals of Ru to vacant 4d \ce{e_g} orbitals on the metal center. Similarly, the C features with the C1 and C2 peaks along a CIE of (experimental) 2843.5 eV correspond to transitions from Ru 2\ce{p_{3/2}} orbitals to unoccupied orbitals of mixed Ru 4d-CN $\pi^*$ character. The splittings between B1(C1) and B2(C2) along the energy transfer axis reflect to emission of a photon due the relaxation of either Ru 3\ce{d_{5/2}} or 3\ce{d_{3/2}} electrons to fill the 2\ce{p_{3/2}} hole. Therefore, the energy separation between these subpeaks provides a direct measure of the splitting induced by 3d spin–orbit coupling (SOC). Experimentally, the 3d SOC was found to be 4.1 eV, whereas its theoretical two- and four-component TDDFT predictions of 4.3 eV for B-splitting and 4.4 eV for C-splitting are slightly overestimated. In the incident energy direction, the experimental energy difference between the B and C features is 1.8 eV and increases to 2.3 eV in our theoretical prediction. Along a diagonal cut (constant emission energy), lie either the B1 and C1 features or the B2 and C2 features. The energy splitting between the 1 and 2 peaks in a given feature, i.e., B2-B1, corresponds to the 3d spin-orbit coupling.

\begin{figure}[H]
    \centering
    \includegraphics[width=1.0\linewidth]{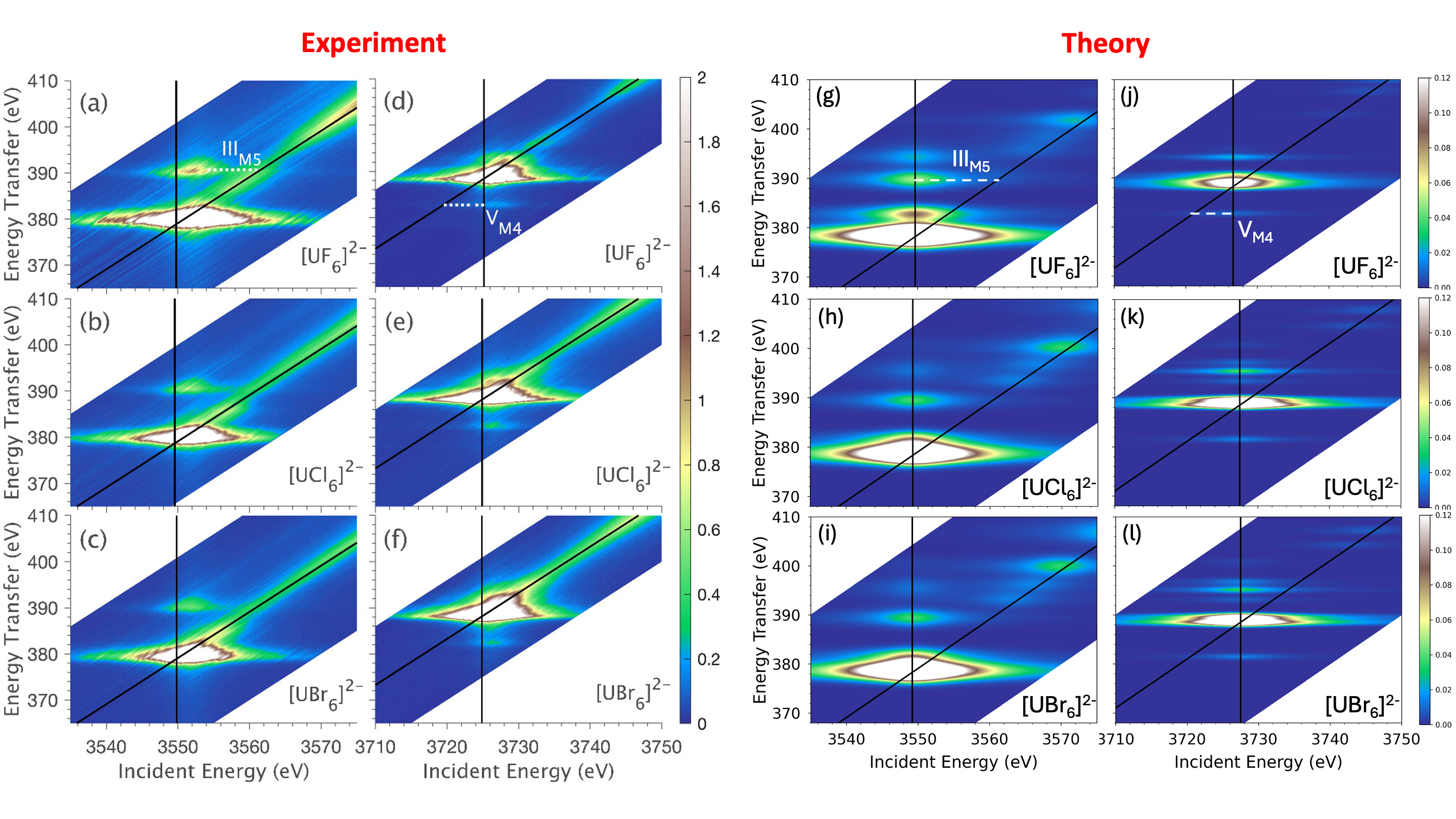}
    \caption{3d4f RIXS maps of \ce{[UX6]^{2-}} measured experimentally at the \ce{M5}-edge (a–c) and \ce{M4}-edge (d–f) in Ref.~\citenum{Burrow2024}, and computed theoretically at the same edges (g–l) in this work. Theoretical values were obtained at the relativistic two-component amfX2C TDDFT level using the PBE0-60HF functional, with the broadening parameters $\Gamma = \unit[8.8]{eV}$, $\gamma = \unit[2.5]{eV}$ (\ce{M5}), and $\gamma = \unit[1.0]{eV}$ (\ce{M4}) as discussed in Eq.~\eqref{eq:RIXS}. The diagonal black lines indicate the cuts taken to obtain HERFD (high-energy resolution fluorescence detection) spectra, while the vertical black lines indicate resonant RXES (X-ray emission spectroscopy) cuts. The white dashed lines show how the intensities from features \ce{III_{M5}} and \ce{V_{M4}} contribute to both the HERFD and the RXES cuts.}
    \label{fig:RIXS:UX6}
\end{figure}

\textbf{\ce{[U^{IV}X6]^{2-}} (X = F, Cl, Br):} In all complexes, the experimental \ce{M5}-edge RIXS is characterized by two intense features, both centered around 3550 eV incident energy and attributed to the resonant 3\ce{d_5/2} $\to$ 5f absorption (Figure \ref{fig:RIXS:UX6} a-c). The second weaker feature, labeled as $\text{III}_{\text{M5}}$, is separated from the primary feature by about 10 eV in the energy transfer axis, due to spin-orbit splitting of the 4f orbitals. The intensity of the $\text{III}_{\text{M5}}$ feature decreases from F to Br. In addition, the broad profile of this feature is so broad that it extends to the diagonal HERFD cut, and thus contributes to the \ce{M5}-edge HERFD spectra. The \ce{M4}-edge RIXS (Figure \ref{fig:RIXS:UX6} d-f), on the other hand, is characterized by a single strong 3\ce{d3/2} $\to$ 5f absorption feature, and a low-intensity satellite labeled as $\text{V}_{\text{M4}}$, both around 3725 eV incident energy.

In general, our theoretical RIXS maps reproduce the experimental observations, as well as the ligand-field DFT planes reported in Figure 8 of Ref.~\citenum{Burrow2024}, with the positions and shapes of the primary spectral features correctly captured. Most importantly, the simulations are able to capture the \ce{V_{M4}} pre-edge feature around 380 eV in the energy transfer axis --- a direct reporter of the covalency of the metal-ligand bonds in these complexes~\cite{Burrow2024}. Interestingly, no shift was applied in neither the incident energy nor the energy transfer axis to align the theoretical and experimental spectra, owing to the good performance of TDDFT with the modified PBE0-60HF functional for these systems. Furthermore, the theoretical results correctly describe the decreasing area of the \ce{M5} main feature and the increasing area of the \ce{M4} main feature from F to Br.
\highlight{The correct reproduction of these trends indicates that the TDDFT approach captures not only the energies but also the relative intensities of the spectral transitions contributing to the \ce{M4} and \ce{M5} features.}
The main features at both edges, but particularly \ce{M4} exhibit certain asymmetry that is only in part reproduced in the calculations. This may be attributed to the fact that our current methodology does not incorporate multiplet effects.~\cite{Butorin20203d}

\begin{figure}[H]
    \centering
    \includegraphics[width=1.0\linewidth]{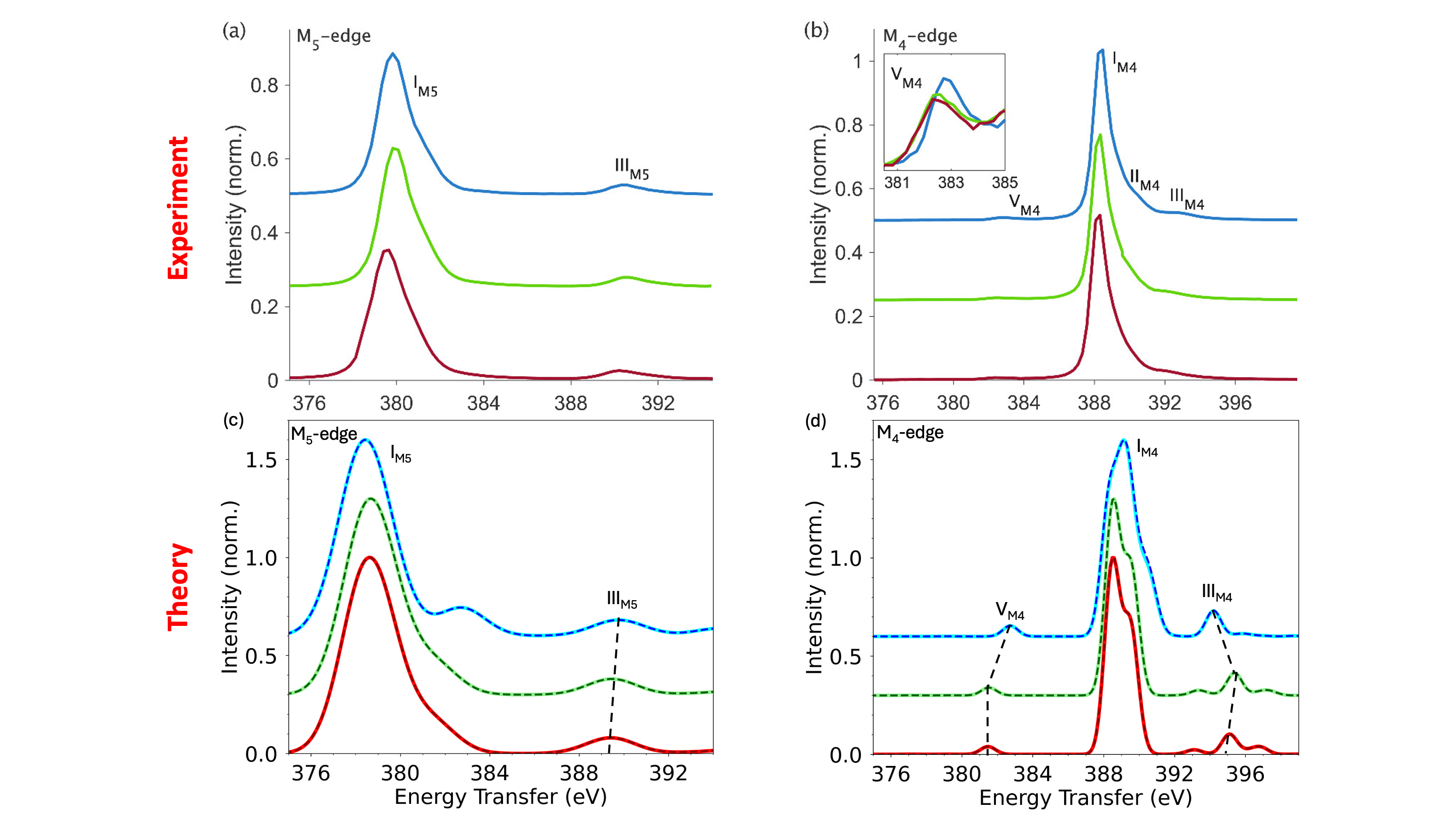}
    \caption{RXES spectra measured experimentally (a--b) in Ref.~\citenum{Burrow2024} and computed theoretically (c--d) in this work at the \ce{M5}-edge (a, c; left) and \ce{M4}-edge (b, d; right) for \ce{[UF6]^{2-}} (blue, top line), \ce{[UCl6]^{2-}} (green, middle line), and \ce{[UBr6]^{2-}} (red, bottom line). The RXES spectra were obtained as vertical cuts of the RIXS maps shown in Fig.~\ref{fig:RIXS:UX6}. In experimental spectra, inset plots highlight the identified satellite features. Theoretical spectra were computed at the relativistic four-component (solid lines) and two-component amfX2C (dashed lines) TDDFT levels using the PBE0-60HF functional.}
    \label{fig:RXES:UX6}
\end{figure}

In the RXES spectra presented in Figure~\ref{fig:RXES:UX6}, the main lines are also well reproduced, specially at the \ce{M4} edge, where the \ce{I_{M4}} signal assigned to the dominant \ce{4f_{5/2}} $\to$ \ce{3d_{3/2}} transition appearing near 388 eV and the \ce{III_{M4}} feature around 394 eV. Here, the experimental red shift observed for the \ce{V_{M4}} feature across the \ce{[UF6]^{2-}}, \ce{[UCl6]^{2-}}, \ce{[UBr6]^{2-}} series is clearly observed in the theoretical spectra. At the \ce{M5} edge, we are also able to capture the main \ce{4f_{7/2}} $\to$ \ce{3d_{5/2}} (\ce{I_{M5}}) and \ce{4f_{5/2}} $\to$ \ce{3d_{5/2}} (\ce{III_{M5}}) lines reported experimentally at 380 and 390 eV, respectively, albeit with a small red shift. \highlight{The weak feature predicted theoretically at approximately 383.5 eV in the \ce{M5}-edge spectrum of \ce{[UF6]^{2-}} corresponds to a low-intensity shoulder observed experimentally on the high-energy side of the main peak; however, its energy and intensity are overestimated by the present TDDFT calculations, most likely due to limitations of the approximate exchange--correlation functional and kernel.}

\begin{figure}[H]
    \centering
    \includegraphics[width=1.0\linewidth]{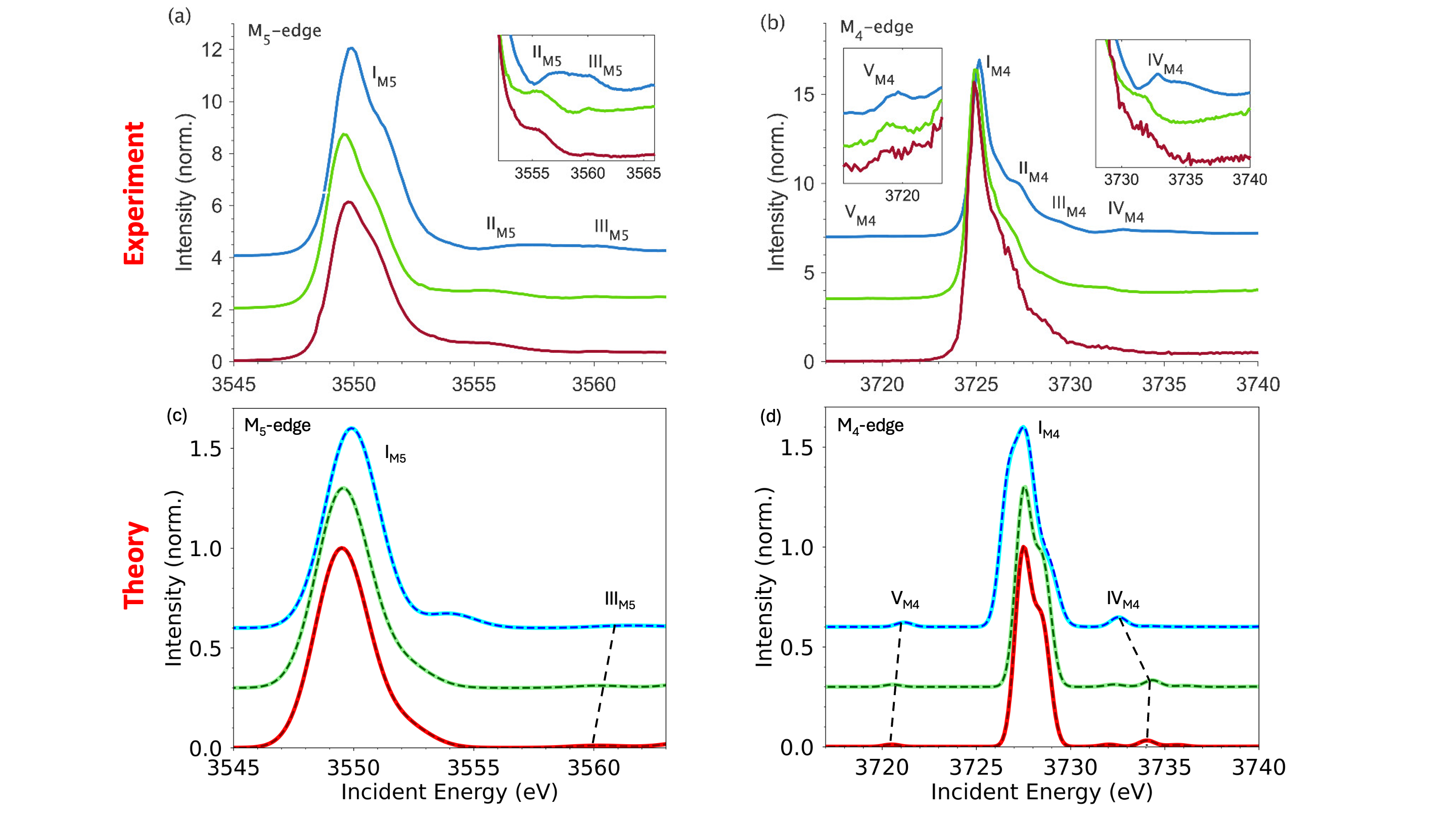}
    \caption{HERFD spectra measured experimentally (a--b) in Ref.~\citenum{Burrow2024} and computed theoretically (c--d) in this work at the \ce{M5}-edge (a, c; left) and \ce{M4}-edge (b, d; right) for \ce{[UF6]^{2-}} (blue, top line), \ce{[UCl6]^{2-}} (green, middle line), and \ce{[UBr6]^{2-}} (red, bottom line). The HERFD spectra were obtained as diagonal cuts of the RIXS maps shown in Fig.~\ref{fig:RIXS:UX6}. In the experimental spectra, inset plots highlight the identified satellite features. Theoretical spectra were computed at the relativistic four-component (solid lines) and two-component amfX2C (dashed lines) TDDFT levels using the PBE0-60HF functional.}
    \label{fig:HERFD:UX6}
\end{figure}

The theoretical HERFD spectra, taken at the maxima of the emission lines (\ce{4f_{5/2}} $\to$ \ce{3d_{3/2}} and
\ce{4f_{7/2}} $\to$ \ce{3d_{5/2}}) and shown in Figure \ref{fig:HERFD:UX6}, also achieve correct relative positions and trends. In particular, we note the good agreement for the \ce{I_{M5}} and \ce{III_{M5}} (Fig. \ref{fig:HERFD:UX6} a and c), and the \ce{I_{M4}}, \ce{IV_{M4}}, and \ce{V_{M4}} (Fig. \ref{fig:HERFD:UX6} b and d) features.

\section{Conclusion}

In this work, we have presented the development and implementation of a relativistic quantum chemical methodology for the calculation of resonant-inelastic X-ray scattering (RIXS) spectra represented as two-dimensional maps, as well as the related high-energy-resolution fluorescence detection (HERFD) and resonant X-ray emission spectra (RXES) from cuts of the RIXS maps. The approach is based on relativistic four-component Dirac--Coulomb and two-component amfX2C Hamiltonians, while the electronic structure and response properties are described within (linear-response time-dependent) density functional theory. Efficient evaluation of transition moments between excited states is achieved by the pseudo-wavefunction formalism. The methodology has been implemented in the quantum chemical program \ReSpect{} and benchmarked on the ruthenium \ce{L3}-edge in \ce{[Ru^{II}(CN)6]^{4-}} and the uranium \ce{M4}- and \ce{M5}-edges in hexahalide complexes, \ce{[U^{IV}X6]^{2-}} (X = F, Cl, or Br).

In the studied systems, characteristic RIXS spectral features, such as positions, lineshapes, and intensities, are governed by the spin–orbit coupling of 2p, 3d, and 4f orbitals. The computed RIXS, RXES, and HERFD spectra successfully reproduce these features and the corresponding experimental trends, in particular the signatures of spin--orbit splitting as well as characteristics associated with covalency. Furthermore, the two-component amfX2C approach reproduces the reference four-component results with near-perfect accuracy while providing an almost order-of-magnitude reduction in computational cost.

Building on our previous works on X-ray absorption, circular dichroism, and transient absorption spectroscopies, this study constitutes a further step in the development of two- and four-component relativistic quantum chemical methods for core spectroscopies in the \ReSpect{} program. It thus expands the available toolbox for researchers seeking efficient yet accurate approaches to modeling high-energy spectra, particularly for systems containing heavy elements, where a first-principles treatment of relativistic effects is essential. More broadly, the demonstrated capability of relativistic quantum chemical methods to reliably describe diverse molecular properties for elements across the entire periodic table underscores the enduring significance of unifying quantum mechanics with special relativity, an idea that traces back to Schr{\"o}dinger's original aspirations.

\begin{acknowledgement}
We acknowledge support from the Research Council of Norway through its Centres of Excellence scheme (project No.~262695) and research grant No.~315822. M.R. and L.K. acknowledge funding from the EU NextGenerationEU through the Recovery and Resilience Plan for Slovakia under the project No.~09I05-03-V02-00034 and No.~09I01-03-V04-00096, as well as from the Slovak Research and Development Agency (grant No.~APVV-22-0488) and VEGA (grant No.~1/0670/24). M.~A.~D. and D.~R.~N. acknowledge support by the National Science Foundation under the CAREER grant No. CHE-2337902. The calculations were performed using resources provided by the High-Performance Computing Facility of the University of Memphis and by Sigma2, the National Infrastructure for High Performance Computing and Data Storage in Norway (grant No.~NN14654K)..
\end{acknowledgement}

\bibliography{manuscript}

\newpage
\begin{figure}
    \centering
    \includegraphics[width=1.00\linewidth]{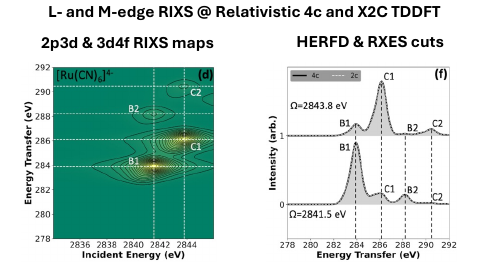}
    \caption{TOC Graphic}
\end{figure}

\end{document}